\documentclass[12pt,fleqn]{article}
\usepackage[dvips]{color}
\usepackage[dvips]{graphicx}

\begin{document}
\title{Universal tunneling time for all fields}

\author{G. Nimtz$^{1,2}$ and A. A. Stahlhofen$^1$\\
$^1$Institut f\"ur Integrierte Naturwissenschaften, Universit\"at
Koblenz\\
$^2$II. Physikalisches Institut, Universit\"at zu K\"oln}

\date{\today}
\maketitle

\begin{abstract}

Tunneling is probably the most important physical process. The
observation that particles surmount a high mountain in spite of
the fact that they don't have the necessary energy can not be
explained by classical physics. However, this so called tunneling
became allowed by the theory of quantum mechanics. Experimental
tunneling studies with different photonic barriers from microwave
frequencies up to ultraviolet frequencies pointed toward  a
universal tunneling time~\cite{Haibel,Esposito}. The observed
results and calculations have shown that the tunneling time of
opaque photonic barriers (for instance optical mirrors) equals
approximately the reciprocal frequency of the electromagnetic wave
 in question. The tunneling process is described by virtual
photons~\cite{Stahlhofen}. Virtual particles like photons or
electrons are not observable. However, from the theoretical point
of view, they represent necessary intermediate states between
observable real states. In the case of tunneling there is a
virtual particle between the incident and the transmitted
particle. Tunneling modes have a purely imaginary wave number.
They represent solutions of the Schr\"odinger equation and of the
classical Helmholtz equation. The most prominent example  of the
occurrence of tunneling modes in optics is frustrated total
internal reflection (FTIR) at double prisms. In 1949 Sommerfeld
\cite{Sommerfeld} pointed out that this pseudo classical optical
phenomenon represents the analogy of quantum mechanical tunneling.
Recent experimental and theoretical data confirmed the conjecture
that the tunneling process is characterized by a universal
tunneling time independent of the kind of field. Tunneling
proceeds at a time of the order of magnitude of the reciprocal
frequency of the wave.
\end{abstract}

Optical evanescent modes and solutions of quantum mechanical
tunneling have a  purely imaginary wave number. This means that
they do not experience a phase shift in traversing space. The
delay time $\tau$ of a propagating wave packet is given by the
derivative

\begin{eqnarray}
\tau = - d\phi/d\omega,
\end{eqnarray}\\
where $\phi$ is the phase shift of the wave and $\omega$ is the
angular frequency; $\phi$ is given by the real part of the wave
number $k$ times the distance $x$. In the case of evanescent modes
and tunneling solutions the real part of $k$ is zero. (The
imaginary part is often called $\kappa$.) Thus propagation of
evanescent modes appears to take place in zero time. In the case
of particle tunneling Eq.1 is replaced by the corresponding
derivative of the S-matrix with respect to energy.

Despite the fact that the semiconductor tunnel diode has been used
since 1962 the particle barrier penetration time has not been
determined up to now due to parasitic time consuming electronic
interaction effects in a semiconductor. Around 1990 the
mathematical analogy between the Schr\"odinger and the Helmholtz
equations inspired microwave and optical tunneling experiments to
obtain empirical data on the tunneling time. The experiments
revealed superluminal (faster than light) signal and energy
velocities~\cite{Enders,Steinberg,Longhi,Nimtz}.

Several quantum mechanical (QM) and quantum electrodynamical (QED)
calculations predicted that evanescent modes and tunneling of
particles appear to propagate in zero time
\cite{Stahlhofen,Hartman,Carniglia,Ali,Low,Wang}. However, there
arises a scattering time at the barrier entrance in agreement with
the Hartman effect. The latter states that the tunneling time does
not depend on barrier length~\cite{Hartman}. The effect was first
confirmed by microwave experiments~\cite{Nimtz2}.

Superluminal tunneling experiments were carried out first with
microwaves in undersized wave guides by Enders and Nimtz and later
reproduced at microwaves and optical frequencies in photonic band
gap material, i.e. optical mirrors~\cite{Nimtz,Nimtz4}.

The tunneling time is of the order of magnitude of the inverse
frequency of the tunneling wave. This property was observed for
electromagnetic tunneling~\cite{Haibel,Esposito,Nimtz}. Recent
experimental and theoretical results point to a universal
tunneling time for all kinds of fields.

\begin{figure}[htb]
\begin{center}
\includegraphics[width=0.7\textwidth]{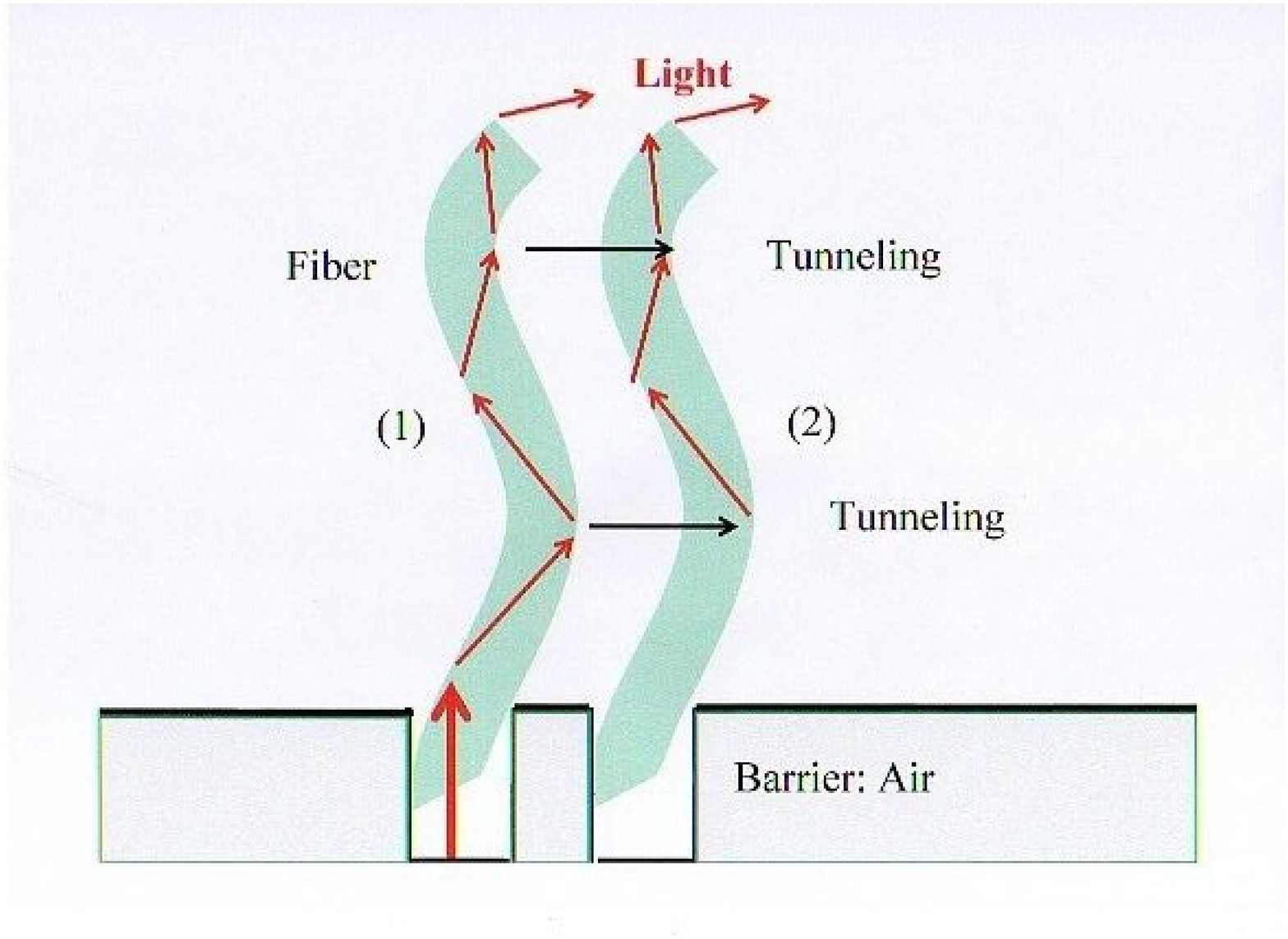}
\end{center}
\caption{Sketch of two fibers (1) and (2). The guided infrared
digital signal beam is totally reflected, however, tunneling takes
placesince the two fibers are approaching each other. A fiber
represents a light valley between the air barrier environment.
\label{fiber}}
\end{figure}


Zero time tunneling and the virtual nature of evanescent modes as
well as the barrier interaction time become most obvious in FTIR
with double prisms as shown in
Refs.~\cite{Stahlhofen,Haibel2,Nimtz3}.

A modern technical device based on FTIR is illustrated in
Fig.\ref{fiber}. The two fibers (1) and (2) are wave guides of
today's digital infrared communication systems~\cite{Longhi}. The
fibers represent light valleys. If they are brought to a narrow
distance of a about a wavelength, more and more intensity of the
infrared signals are tunneling from fiber (1) to fiber (2).
Engineers call such a device a coupler, the inter-fiber distance
determines the coupling ratio between the two guides by the
tunneling process. Here, the barrier is given by an air gap.

In FTIR there is a shift (coined Goos-H\"anchen shift) between the
incident and the reflected beam as already conjectured by Newton
300 years ago.

The length of this shift amounts to about one wavelength
~\cite{Haibel2}. This shift along the boundary of the first prism
corresponds to the universal tunneling time of one oscillation
time ~\cite{Haibel,Esposito}. The shift represents the interaction
time at the boundary in the case that the beam has no normal
incidence to the boundary.

The relationship for the tunneling intensity $I_t$ in the case of
FTIR is

\begin{eqnarray}
I_t(x) & = &  I_0 e^{(i\omega t - 2 \kappa x)} \label{E}\\
\kappa & = & [\frac{\omega^2}{c^2} ((\frac{n_1}{n_2})^2 - 1) sin^2
\theta)]^{1/2}, \label{kappa}
\end{eqnarray}
where $I_0$ is the the signal intensity at the barriers front,
$\kappa$ is the imaginary wave number, $c$, $n_1$ and $n_2$ are
velocity of light, the refractive indices of the air and of the
fibers, respectively, and $\theta$ represents the angle of
incidence above the critical angle of total
reflection~\cite{Haibel2}.

\begin{figure}[hbt]
\vspace*{-0.5cm}
\begin{center}
\includegraphics[width=1\linewidth]{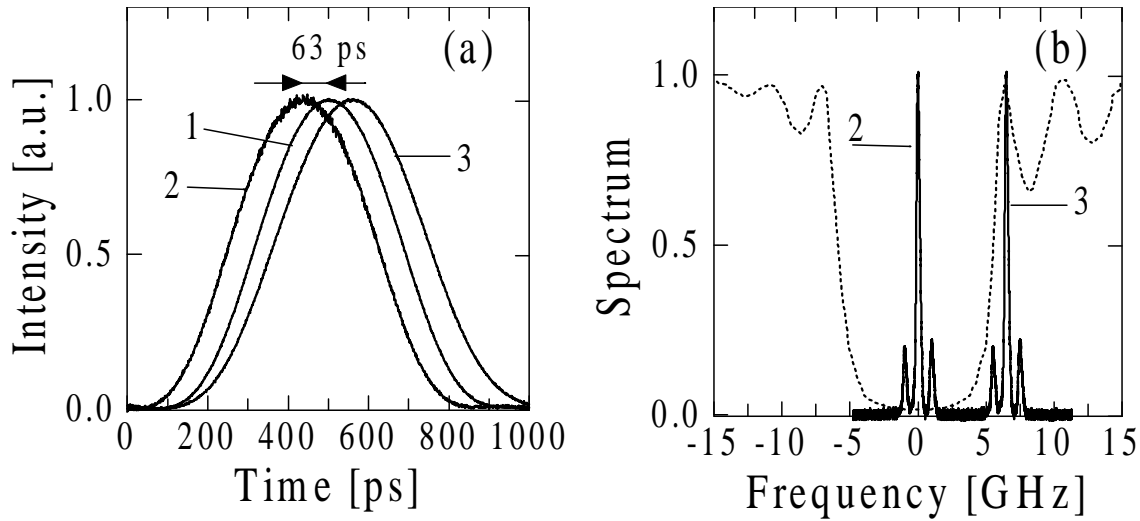}
\vspace*{-0.4cm} \caption{Measured propagation time of three
digital signals \cite{Longhi}. (a) Pulse trace 1 was recorded in
vacuum. Pulse 2 traversed a photonic lattice in the center of the
frequency band gap (see part (b) of the figure), and pulse 3 was
recorded for the pulse travelling through the fiber outside the
forbidden band gap. The photonic lattice was a periodic dielectric
hetero-structure fiber. The IR carrier frequency is 2 $10^{14}$.
The signal speed was 2c.\label{longhi} }
\end{center}
\end{figure}

\begin{figure}[hbt]
\begin{center}
\includegraphics[width=0.70
\linewidth]{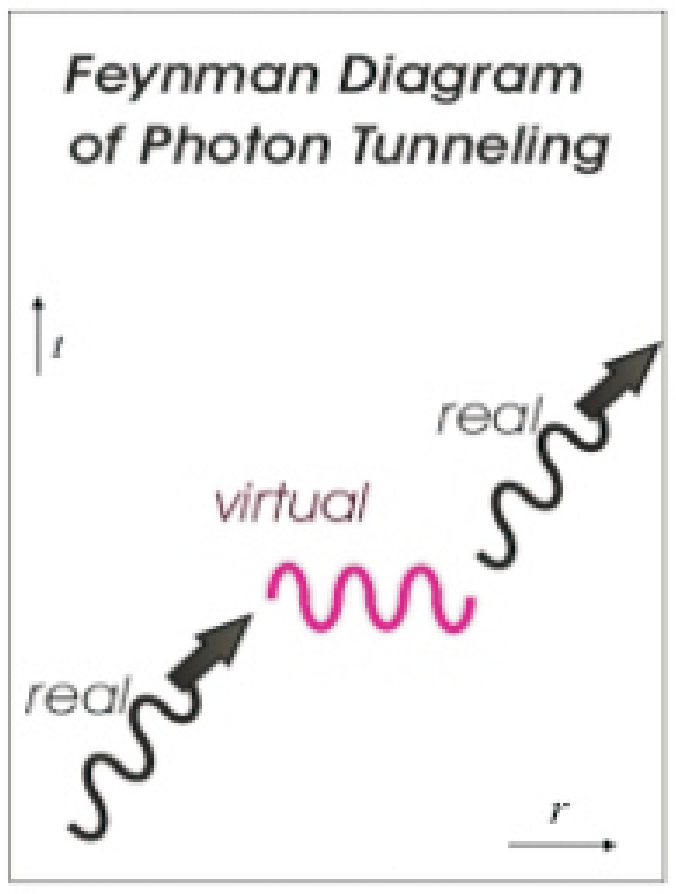}
\end{center}
\caption{Feynman diagram of evanescent mode or photonic tunneling.
Virtual photons connect the observable incoming and transmitted
photons.} \label{Feynman}
\end{figure}

It is interesting that evanescent modes and tunneling particles
are non observables. They can neither be detected in an undersized
waveguide nor in a potential barrier for
instance~\cite{Stahlhofen,Nimtz}. This behavior is in consequence
of the uncertainty relation~\cite{Stahlhofen}.

The universal tunneling time seems to be valid also for sound
waves as measured, for instance, by Yang et al. at a frequency of
1 MHz and by Robertson et al. at 1 kHz in a sound tunneling
experimental set-up~\cite{Yang,Robertson}. Presumably, the virtual
behavior of photons applies for all wave solutions having purely
imaginary wave numbers independent of the kind of field. For
instance electronic tunneling time values measured and discussed
in field emission microscopy and in III-V compound semiconductors
resulting in a time also fitting in the universal tunneling time
scheme~\cite{Pereyra}. Table 1 shows tunneling data measured with
different barriers and fields, where $\tau$ , $\nu$ and $T$ are
the tunneling time, the carrier frequency or a wave packet's
energy E dived by h, and T the oscillation time of the wave. In
Ref.~\cite{Haibel} it was conjectured that the relation also holds
for wave packets with rest mass having in mind the mathematical
analogy between the Helmholtz and the Schr\"odinger equations.
Quantum mechanical studies pointed to this conjecture
\cite{Hartman,Low,Collins}.

An argument has been raised recently that a superluminal signal
speed can never happen (cf. the comment in Ref.~\cite{Steinberg2}.
The superluminal experimental results of digital signals were
interpreted analogous to a train loosing at every station a front
waggon resulting in a faster train speed of the last arriving
waggons. This naive model does not consider that physical signals
are frequency band limited as measured and displayed in
Fig.\ref{longhi} and as explained in Ref.\cite{Nimtz} for
instance. The superluminal signals did not lose any frequency
components (that means no loss of waggons as argued in an improper
analogy ignoring the QM and QED results~\cite{Stahlhofen}). All
components of the superluminal signals studied are transmitted and
correctly detected as shown in Fig.\ref{longhi}. A measurable
pulse reshaping did not take place.

All four properties: i) The universal tunneling time. ii) The
violation of the Einstein energy relation by the imaginary wave
number
\begin{eqnarray}
E^2 = (\hbar k c)^2 + (m_0 c^2)^2
\end{eqnarray}.\\
iii) The zero time spreading (non locality), and iv) The non
observability of evanescent modes, can be explained by virtual
particles in the tunneling process. The Feynman diagram of
photonic tunneling is shown in Fig.\ref{Feynman}.

We are used to expect virtual particles in microscopic interaction
processes, here we have demonstrated examples in the macroscopic
range of a meter in Coulomb and in elastic fields.

\begin{center}
\begin{tabular}{|l|l|c|c|}

\hline \textbf{Table: Tunneling time} &   &    &   \\
\hline
\hline photonic barriers & reference  & $\tau$   &  $T=1/\nu$ \\
\hline
\hline {\it frustrated }   & Haibel/Nimtz & 117\,ps & 120\,ps  \\
\cline{2-4} {\it total reflection}   & Carey et al. & $\approx$ 1\,ps  & 3\,ps \\
\cline{2-4} {\it  at double prism}  & Balcou/Dutriaux  &  30\,fs &  11.3\,fs   \\
\cline{2-4} & Mugnai et al. & 134\,ps & 100\,ps    \\
\hline
\hline {\it photonic lattice} & Steinberg et al. & 2.13\,fs & 2.34\,fs  \\
\cline{2-4} &  Spielmann et al.  &  2.7\,fs  &  2.7\,fs     \\
\cline{2-4} & Nimtz et al.  &  81\,ps  &  115\,ps     \\
\hline
\hline {\it undersized} & Enders/Nimtz & 130\,ps & 115\,ps \\
{\it waveguide}         &                      &               &              \\
\hline
\hline {\it electron tunneling (field  }   & Sekatskii/Letokhov & 6 - 8\,fs & $>$2.43\,fs  \\
\cline{2-4} {\it emission microscopy, }  & Pereyra &  100\,fs &  37.5\,fs       \\
{\it semiconductor AlGaAs-GaAs) }         &                      &               &             \\
\hline
\hline {\it acoustic (phonon) tunneling } & Yang et al. & 0.6 - 1\,$\mu$s &  1\,$\mu$s \\
\cline{2-4}   & Robertson et al. & 0.9\,ms &  1.12\,ms  \\
\hline
\end{tabular}
\end{center}


\end{document}